\documentclass[letterpaper,english,reprint,aps]{revtex4-1}
\usepackage[T1]{fontenc}
\usepackage[latin9]{inputenc}
\usepackage{amsmath}
\usepackage{amssymb}
\usepackage{graphicx}
\usepackage{soul}
\usepackage{colortbl}
\usepackage[colorlinks=true,citecolor=blue,linkcolor=blue]{hyperref}

\usepackage{xcolor}

\makeatletter

\pdfpageheight\paperheight
\pdfpagewidth\paperwidth

\makeatother

\usepackage{babel}
\begin{document}

\title{Quantum generative adversarial networks}

\author{Pierre-Luc Dallaire-Demers}
\email{pierre-luc@xanadu.ai}

\author{Nathan Killoran}

\affiliation{Xanadu, 372 Richmond Street W, Toronto, Ontario M5V 1X6, Canada}

\date{\today}
\begin{abstract}
Quantum machine learning is expected to be one of the first potential general-purpose
applications of near-term quantum devices. A major recent breakthrough in classical machine learning is the notion of generative adversarial training, where the gradients of a discriminator model are used to train a separate generative model. In this work and a companion paper, we extend adversarial training to the quantum domain and
show how to construct generative adversarial networks using quantum circuits. Furthermore, we also show how to compute gradients -- a key element in generative adversarial network training -- using another quantum circuit. We give an example of a simple practical circuit ansatz to parametrize
quantum machine learning models and perform a simple numerical experiment to demonstrate that quantum generative adversarial networks can be trained successfully.
\end{abstract}
\maketitle

\section{Introduction\label{sec:Introduction}}

Deep learning \citep{LeCun2015,Goodfellow2016} is currently transforming the
way we process large-scale complex data with computers. Deep
neural networks are now able to perform image and speech recognition
with accuracies at a similar level to humans \citep{Deng2014}. One of the most exciting recent developments in deep learning is \emph{generative adversarial networks} (GANs) \citep{Goodfellow2014}. These are a class of deep neural
networks which have shown great promise for the task of \emph{generative} machine learning, that is, learning to generate realistic data samples. Despite
the initial difficulties of training these models \citep{Salimans2016},
GANs have quickly found applications in many fields \citep{Creswell2018}, including image generation \citep{zhu2016generative}, super-resolution \citep{ledig2016photo}, image-to-image translation \citep{isola2017image}, generation of 3D objects \citep{choy20163d}, text generation \citep{gulrajani2017improved}, and the generation of synthetic data for chemistry \citep{kadurin2017drugan}, biology \citep{killoran2017}, and physics \citep{de2017learning}.

The goal of GANs is to simultaneously train two functions: a \emph{generator} $G$, and a \emph{discriminator} $D$, through an \emph{adversarial learning} strategy. The goal for the generator is to generate new sample data from some specific domain, such as images, text, or audio. The outputs from the generator should not be completely unstructured; rather, they should be plausible samples that reflect the properties of real-world data (e.g., realistic images or natural language). The goal of the discriminator is to distinguish fake data samples which were created by the generator from those which are real.

The training strategy for GANs is anchored in game theory
and is analogous to the competition between counterfeiters who have
to produce fake currencies and the police who have to design methods
to distinguish increasingly more convincing counterfeits from
the real ones. This game has a Nash equilibrium where the fake coins
become indistinguishable from the real ones and the authorities can
no longer devise a method to discriminate the real currencies from
the generated ones \citep{Goodfellow2014} . 
Interestingly, theoretical proofs regarding the optimal points of adversarial training assume that the generator and discriminator have infinite capacity \citep{Goodfellow2014}, i.e., they can encode arbitrary functions or probability distributions. Yet it is widely believed that classical computers cannot efficiently solve certain hard problems, so these optimal points may be intrinsically out of reach of classical models in many cases of interest. 

Quantum computers \citep{Feynman1982,Nielsen2009} have the potential
to solve problems believed to be beyond the reach of classical computers, such as
factoring large integers \citep{Shor1997}. Realistic near-term quantum devices \citep{Preskill2018}
may be able to speed up difficult optimization and sampling problems, even if the full
power of fault-tolerant devices may not be available for several years.
For instance, variational quantum algorithms \citep{Peruzzo2014,Kandala2017,Moll2017,Guerreschi2017,Dallaire-Demers2018}, such as the variational quantum eigensolver (VQE), 
have been demonstrated with great success in the field of quantum chemistry.
Currently, these ideas and algorithms are being extended to the domain of quantum machine learning \citep{Schuld2014,Arjovsky2015,Romero2017,Romero2017a,Biamonte2017,Cao2017,verdon2017quantum, Otterbach2017,Schuld2018,Schuld2018a,Huggins2018,Farhi2018,Mitarai2018},
which could also benefit from a quantum advantage. Since many machine learning algorithms are naturally robust to noise, this direction is a promising application for near-term imperfect quantum devices.

In this paper, we introduce \emph{QuGANs}, the quantum version of generative adversarial networks. The paper has the following structure. 
In Section \ref{subsec:StructureQGANs}, we generalize the model structure
of classical generative adversarial networks \citep{Goodfellow2014}
to define the quantum mechanical equivalent -- QuGANs -- and provide the cost function for training. 
A key ingredient for GANs is that the discriminator provides a gradient which the generator can use for gradient-based learning.
In Section \ref{subsec:QuantumGradients}, we present a general formalism for computing exact gradients of quantum optimization and machine learning problems using quantum circuits. We then show how these gradients can be combined with a classical optimization routine to train QGANs in Section \ref{subsec:GradientsQGANs}. Finally, we provide an example quantum circuit for both the generator and discriminator in Section \ref{subsec:Ansatz} and show that QuGANs can be trained in practice with a simple proof-of-principle numerical experiment in Section \ref{subsec:Numerics}.

We will explore the practical issues of QuGANs by explicitly constructing quantum circuits for the generator and discriminator and proposing quantum methods for computing the gradients of these circuits. 
A more in-depth theoretical exploration of quantum adversarial learning can be found in the companion paper \cite{lloyd2018}. 

\section{Training QuGANs\label{sec:TrainingQGANs}}

\subsection{The structure of GANs and QuGANs\label{subsec:StructureQGANs}}

\subsubsection{Classical GANs}

We first provide a high-level overview of the GAN architecture \citep{Goodfellow2014}. We suppose that the real-world data comes from some fixed distribution $p_R\left(x\right)$, generated by some (potentially complex and unknown) process $R$. The generator -- parameterized by a vector of real-valued parameters $\vec{\theta}_{G}$ -- takes as input an unstructured random variable $z$ (typically drawn from a normal or uniform distribution). G transforms this noise 
source into data samples $x=G(\vec{\theta}_{G},z)$, creating the generator distribution $p_G(x)$.
In the ideal case of a perfectly trained generator $G$, the discriminator would not be able to decide whether a
given sample $x$ came from $p_G\left(x\right)$
or from $p_R\left(x\right)$. Therefore, the task of training
$G$ corresponds to the task of maximizing the probability that $D$
misclassifies a generated sample as an element of the real data. 
On the other hand, the discriminator -- parameterized by a vector of real-valued parameters $\vec{\theta}_{D}$ -- takes as input either real data examples $x\sim p_R(x)$ or fake data samples $x\sim p_G(x)$. D's goal is to discriminate between these two classes, outputting a binary random variable. Training $D$ thus corresponds to maximizing the probability of successfully classifying real data, while minimizing the probability of misclassifying fake data.

We will formalize QuGANs as a quantum generalization of conditional
GANs \citep{Mirza2014}. Conditional GANs generate samples from a \emph{conditional} distribution $p(x|\lambda)$ (conditioned on labels $\lambda$), rather than the unconditional distribution $p(x)$ of vanilla GANs. Conditional GANs reduce to vanilla GANs in the case where the label is uninformative about the data, i.e., $p(x|\lambda)=p(x)$ for all $x$ and $\lambda$.
A possible motivation for using the conditional approach comes
from performing quantum chemistry calculations on quantum computers. 
For example,
one could have a list of VQE state preparations for molecules,
labeled by their physical properties. A well-trained QuGAN could produce
new molecular states which also have the same properties but were
not in the original dataset. In another context, a QuGAN could be used
to compress time evolution gate sequences \citep{Kivlichan2018} for
different time steps to use in larger quantum simulations.

\subsubsection{Quantum GANs}
\begin{figure}
\centering{}\includegraphics[width=0.98\columnwidth]{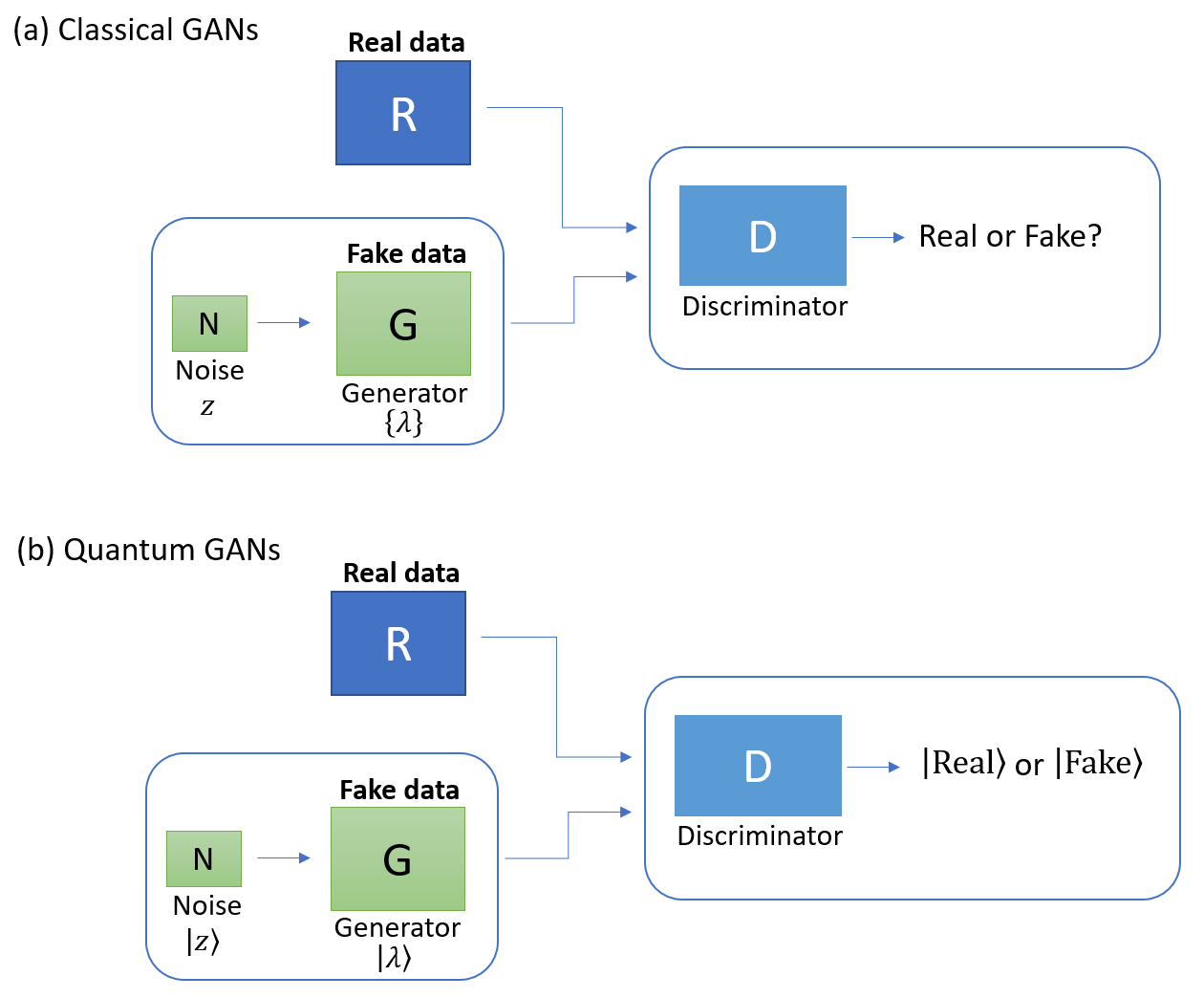}\caption{
In (a), we show the building blocks of classical GANs. A discriminator must determine whether the samples it is given are produced by a real source $R$ or a generator $G\left(z\right)$ equipped with a source of noise $z$. In (b), a quantum discriminator must decide whether the quantum state it receives at its input come from a real quantum data source $R$ or a quantum generator $G\left(\left|z\right\rangle\right)$ with a quantum noise vector $\left|z\right\rangle$. The discriminator yields its output as a quantum state $\left|\mathrm{Real}\right\rangle$ or $\left|\mathrm{Fake}\right\rangle$.
\label{fig:ClassicalQuantumGANs}}
\end{figure}

We will now generalize these ideas to the quantum setting. 
In Figure \ref{fig:ClassicalQuantumGANs}, we highlight the structural similarities of classical and quantum GANs.
For the quantum case, suppose we are given a data source $R$ which, given a label $\left|\lambda\right\rangle $, outputs a density matrix
$\rho_{\lambda}^{R}$ into a register containing $n$ subsystems, i.e.,
\begin{equation}
R\left(\left|\lambda\right\rangle \right)=\rho_{\lambda}^{R}.\label{eq:SourceSpecification}
\end{equation}
The general aim of training a GAN is to find a generator $G$ which
mimics the real data source $R$. In the quantum case, we define
$G$ to be a variational quantum circuit whose gates are parametrized
by a vector $\vec{\theta}_{G}$. The generator takes as input the label $\left| \lambda \right\rangle$ and an additional state $\left| z \right\rangle$, and produces a quantum state, 
\begin{equation}
G(\vec{\theta}_{G},\left|\lambda,z\right\rangle )=\rho_{\lambda}^{G}(\vec{\theta}_{G},z),\label{eq:GeneratorSpecification}
\end{equation}
where $\rho_{\lambda}^{G}$ is output on a register containing $n$ subsystems, similar to the real data. 

The role of the extra input state $\left| z \right\rangle$ is two-fold. On one hand, it can be seen as a source of unstructured noise which provides entropy within the distribution of generated data. For instance, we could have a generator which is unitary, producing a fixed state $\rho_\lambda^G(\vec{\theta}_G, z) = \left| \psi_\lambda(z) \right\rangle \left\langle \psi_\lambda(z) \right|$ for each $\lambda$ and $\left| z \right\rangle$. By allowing the input $\left| z \right\rangle$ to randomly fluctuate, we can create more than one output state for each label. On the other hand, the variable $\left| z \right\rangle$ can serve as a control for the generator. By tuning $\left| z \right\rangle$, we can transform the output state prepared by the generator, varying properties of the generated data which are not captured by the labels $\lambda$. During training, the generator should learn to encode the most important intra-label factors of variation with $\left| z \right\rangle$. While the first role could have been accomplished via coupling the generator to a bath, the second role requires $\left| z \right\rangle$ to be under our control, even if we endow it with no explicit structure during training.

As in the classical case, the training signal of the generator
is provided by a discriminator $D$, made up of separate quantum circuit parametrized
by a vector $\vec{\theta}_{D}$. The task of $D$ is to determine
whether a given input state was created by the real data source $R$
or the generator $G$, whereas the task of $G$ is to fool $D$ into
accepting its output as being real. If the input was created by $R$, then
$D$ should output $\left|\mathrm{real}\right\rangle $ in its 
output register, otherwise it should output $\left|\mathrm{fake}\right\rangle $.
The discriminator is also allowed to do operations on an internal workspace.
In order to force $G$ to respect the supplied labels, 
the discriminator is also given an unaltered copy of the label
$\left|\lambda\right\rangle $. 

The optimization objective for QuGAN training can be formalized as the adversarial task $\min_{\vec{\theta}_{G}}\max_{\vec{\theta}_{D}}V(\vec{\theta}_{D},\vec{\theta}_{G})$, or:
\begin{multline}
\min_{\vec{\theta}_{G}}\max_{\vec{\theta}_{D}}\frac{1}{\varLambda}\sum_{\lambda=1}^{\Lambda}\Pr\Biggl(\left(D(\vec{\theta}_{D},\left|\lambda\right\rangle ,R(\left|\lambda\right\rangle ))=\left|\mathrm{real}\right\rangle \right)\\
\cap\left(D(\vec{\theta}_{D},\left|\lambda\right\rangle ,G(\vec{\theta}_{G},\left|\lambda,z\right\rangle ))=\left|\mathrm{fake}\right\rangle \right)\Biggr).\label{eq:OptimizationTask}
\end{multline}
For classical GANs, the optimization task is traditionally defined with log-likelihood functions but it is more convenient to define a cost function linear in the output probabilities of $D$ in the quantum case since we want to optimize a function which is linear in some expectation value. Since the logarithmic function is convex, the optimal points are the same. Finally, for simplicity, the formula above assumes that the labels are countable, with cardinality $\Lambda$, though this could be relaxed.

\begin{figure}
\centering{}\includegraphics[width=\columnwidth]{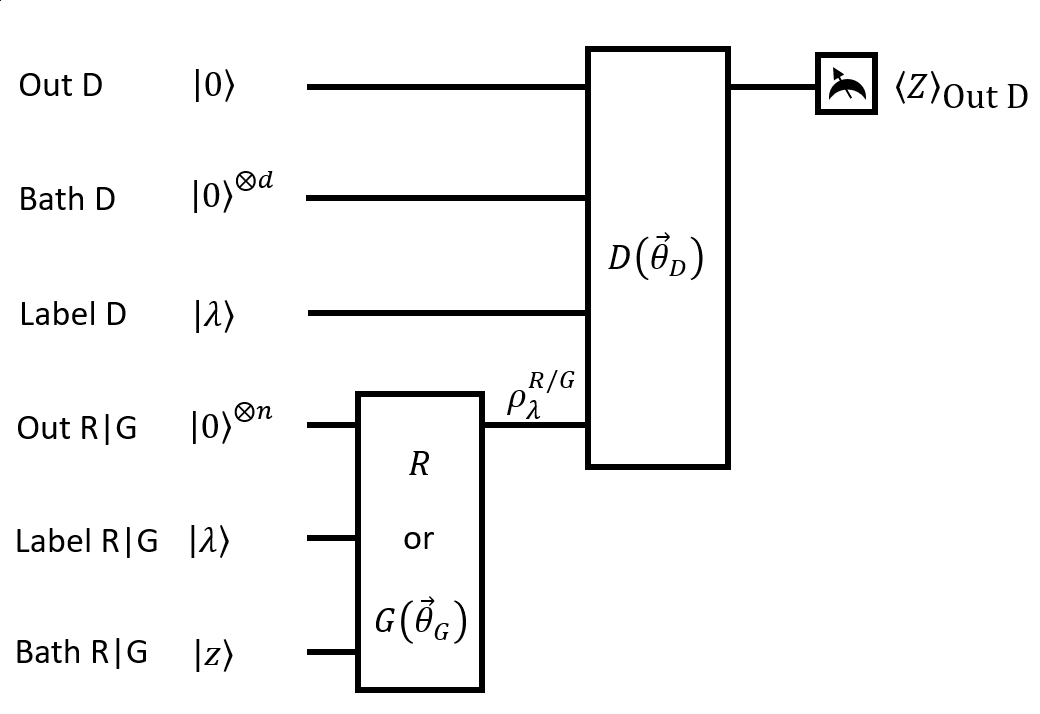}\caption{The general structure of QuGANs. The real source $R$ or the parametrized
generator $G(\vec{\theta}_{G})$ is applied on an initial
state $\left|0,\lambda,z\right\rangle $ respectively defined on the
\textbf{Label R|G}, \textbf{Out R|G} and \textbf{Bath R|G} registers.
The discriminator $D(\vec{\theta}_{D})$ uses the information
$\rho_{\lambda}^{R/G}$ from the source and an initial resource state
$\left|0,0,\lambda\right\rangle $ defined on the \textbf{Out D},\textbf{
Bath D} and \textbf{Label D} registers. $D$ outputs its answer $\left|\mathrm{real}\right\rangle $
or $\left|\mathrm{fake}\right\rangle $ in the \textbf{Out D} register.
The expectation value $\left\langle Z\right\rangle _{\mathrm{Out\:D}}$
is proportional to the probability that $D$ outputs $\left|\mathrm{real}\right\rangle $.
\label{fig:StructureQGAN}}
\end{figure}

The heuristic of the algorithm is illustrated in Figure \ref{fig:StructureQGAN},
where the quantum circuit is divided into 6 operationally defined
registers. The real source $R$ and the generator $G$ are given a
label $\left|\lambda\right\rangle $ in the $s$-subsystem register \textbf{Label
R|G}, an initial blank state $\left|0\right\rangle ^{\otimes n}$
on the $n$-subsystem register \textbf{Out R|G} and a noise vector $\left|z\right\rangle $
on the $m$-subsystem register \textbf{Bath R|G}. In
this work, we assume that $R$ is a purified unitary operation on
$s+n+m$ subsystems. In general, the real source may be a physical device
entangled with an unknown number of environmental degrees of freedom $m'$, with
$m'\neq m$. With no loss of generality, we can assume that the \textbf{Bath R|G} register is initialized in the reference state $\left|0\right\rangle ^{\otimes m}$ when the source is $R$ as the entropy can be provided by the environment. We assume that the discriminator does not have access to
the \textbf{Bath R|G} register. 

$D$ outputs its answer $\left|\mathrm{real}\right\rangle $
or $\left|\mathrm{fake}\right\rangle $ on the register \textbf{Out
D}. It is given the state of the source through register
\textbf{Out R|G}. The workspace of the discriminator
is defined on the $d$-subsystem register \textbf{Bath D} and a reference
copy of the label $\left|\lambda\right\rangle $ is fed through the
$s$-subsystem register \textbf{Label D}. Finally, the expectation value
of the operator 
\begin{equation}
Z\equiv\left|\mathrm{real}\right\rangle \left\langle \mathrm{real}\right|-\left|\mathrm{fake}\right\rangle \left\langle \mathrm{fake}\right|
\label{eq:DecisionDipole}
\end{equation}
on the \textbf{Out D} register is proportional to $\Pr\left(D(\vec{\theta}_{D},\left|\lambda\right\rangle ,R(\left|\lambda\right\rangle))=\left|\mathrm{real}\right\rangle \right)$
and can be used to define the optimization problem \eqref{eq:OptimizationTask}
in a fully quantum mechanical setting.

\subsubsection{The quantum cost function\label{subsec:QuantumCostFunction}}

\begin{figure*}
\centering{}\includegraphics[width=0.8\textwidth]{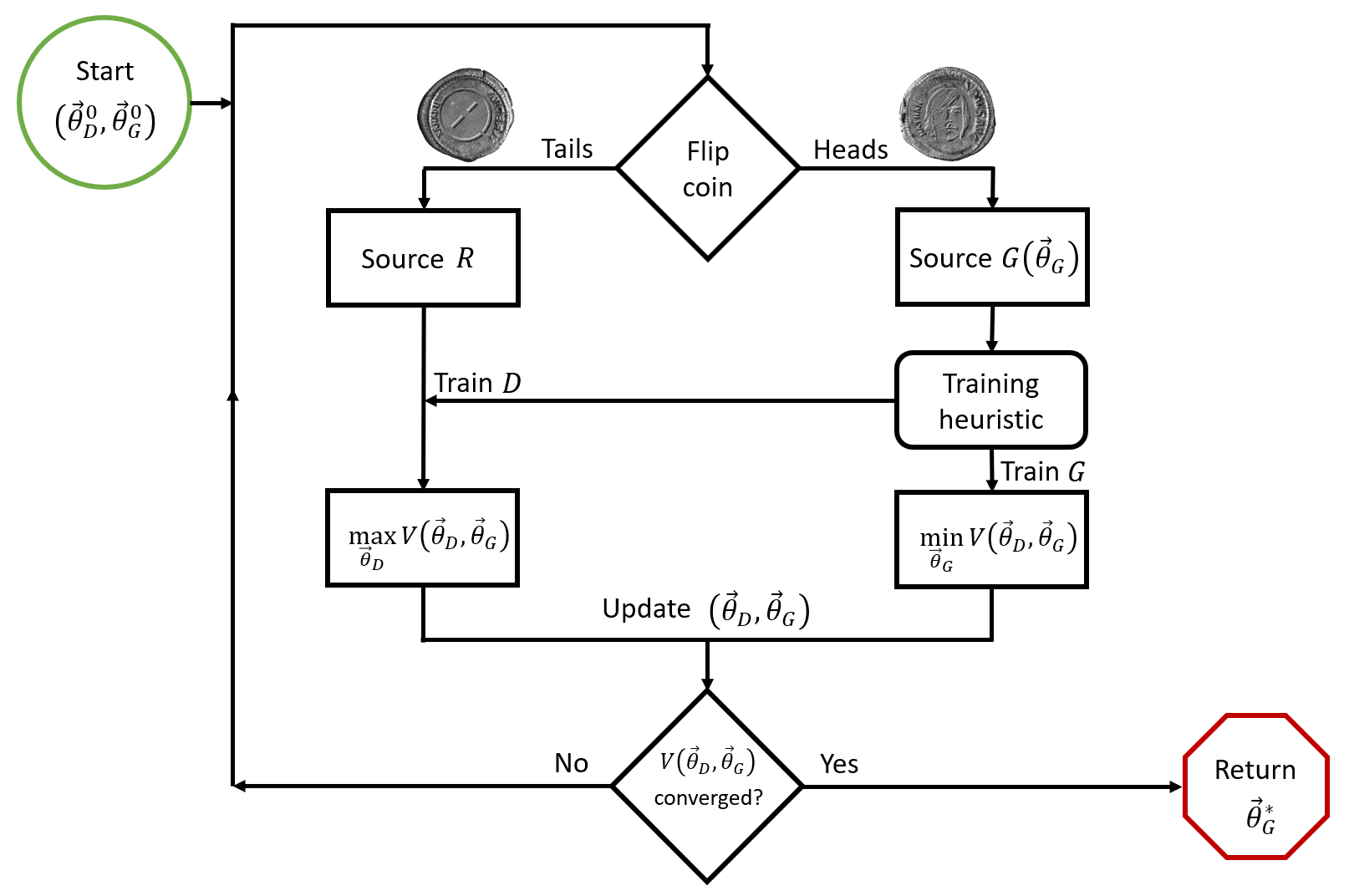}\caption{We illustrate the algorithmic flow of the training of a QuGAN (see text for details). \label{fig:TrainingAlgo}}
\end{figure*}

We will follow the flow of the training process as illustrated in
Figure \ref{fig:TrainingAlgo} to rewrite and analyze the quantum
version of the cost function \eqref{eq:OptimizationTask}. At
the beginning of the algorithm, the discriminator and the generator
are respectively initialized by the (arbitrary) parameters $(\vec{\theta}_{D}^{0},\vec{\theta}_{G}^{0})$.
The quantum computer of Figure \ref{fig:StructureQGAN} is initialized
in the state 
\begin{equation}
\rho_{\lambda}^{0}\left(z\right)=\left(\left|0\right\rangle \left\langle 0\right|\right)^{\otimes d+1}\otimes\left|\lambda\right\rangle \left\langle \lambda\right|\otimes\left(\left|0\right\rangle \left\langle 0\right|\right)^{\otimes n}\otimes\left|\lambda\right\rangle \left\langle \lambda\right|\otimes\left|z\right\rangle \left\langle z\right|.\label{eq:InitialState}
\end{equation}
If only either $R$ or $G$ were systematically fed into $D$, the optimal
strategy of the latter to maximize the cost function \eqref{eq:OptimizationTask} would
be to trivially output a constant answer, which is not desirable. In
order to make sure that $D$ cannot rely on the statistics of the
choice of the source to determine its answer, the choice of $R$ or $G$ can be made
by the toss of a fair coin. The unitary operations corresponding to
the sources $R$ and $G(\vec{\theta}_{G})$ acting on the
whole quantum computer have the respective form 
\begin{equation}
\begin{array}{rcl}
U_{R} & = & I^{\otimes\left(1+d+s\right)}\otimes R,\\
\\
U_{G}(\vec{\theta}_{G}) & = & I^{\otimes\left(1+d+s\right)}\otimes G(\vec{\theta}_{G}).
\end{array}\label{eq:UnitarySources}
\end{equation}
After the chosen source has been applied, the quantum computer is
in the corresponding state
\begin{equation}
\begin{array}{rcl}
\rho_{\lambda}^{R} & = & U_{R}\rho_{\lambda}^{0}\left(0\right)U_{R}^{\dagger},\\
\\
\rho_{\lambda}^{G}(\vec{\theta}_{G},z) & = & U_{G}(\vec{\theta}_{G})\rho_{\lambda}^{0}\left(z\right)U_{G}^{\dagger}(\vec{\theta}_{G}).
\end{array}\label{eq:SourceStates}
\end{equation}
The unitary operation defining the discriminator $D(\vec{\theta}_{D})$
has the form 
\begin{equation}
U_{D}(\vec{\theta}_{D})=D(\vec{\theta}_{D})\otimes I^{\otimes m}\label{eq:UnitaryDiscriminator}
\end{equation}
such that the state of the quantum computer when $U_{D}(\vec{\theta}_{D})$
follows $U_{R}$ is given by
\begin{equation}
\rho_{\lambda}^{DR}(\vec{\theta}_{D})=U_{D}(\vec{\theta}_{D})\rho_{\lambda}^{R}U_{D}^{\dagger}(\vec{\theta}_{D})\label{eq:StateDR}
\end{equation}
and the state when $U_{D}(\vec{\theta}_{D})$ is applied
after $U_{G}(\vec{\theta}_{G})$ is
\begin{equation}
\rho_{\lambda}^{DG}(\vec{\theta}_{D},\vec{\theta}_{G},z)=U_{D}(\vec{\theta}_{D})\rho_{\lambda}^{G}(\vec{\theta}_{G},z)U_{D}^{\dagger}(\vec{\theta}_{D}).\label{eq:StateDG}
\end{equation}
The cost function \eqref{eq:OptimizationTask} can then be written
in the quantum formalism as
\begin{multline}
V(\vec{\theta}_{D},\vec{\theta}_{G})=\frac{1}{2}+\frac{1}{2\Lambda}\sum_{\lambda=1}^{\Lambda}\Biggl(\cos^{2}\left(\phi\right)\mathrm{tr}\left(Z\rho_{\lambda}^{DR}(\vec{\theta}_{D})\right)\\
-\sin^{2}\left(\phi\right)\mathrm{tr}\left(Z\rho_{\lambda}^{DG}(\vec{\theta}_{D},\vec{\theta}_{G},z)\right)\Biggr)\label{eq:QuantumCostNoCoin}
\end{multline}
where both parts depend on $\vec{\theta}_{D}$ and only the second
part depends on $\vec{\theta}_{G}$, as in the classical case \citep{Goodfellow2014}.
Here the angle $\phi$ parametrizes the bias of the coin used in Figure
\ref{fig:TrainingAlgo} since the probability that $R$ or $G$ is used
as a source is not explicitly constrained in \eqref{eq:OptimizationTask}.
Assuming a fair coin $\phi=\frac{\pi}{4}$, the quantum optimization
problem has the final form
\begin{equation}
\min_{\vec{\theta}_{G}}\max_{\vec{\theta}_{D}}\frac{1}{2}+\frac{1}{4\Lambda}\sum_{\lambda=1}^{\Lambda}\mathrm{tr}\Biggl(\left(\rho_{\lambda}^{DR}(\vec{\theta}_{D})-\rho_{\lambda}^{DG}(\vec{\theta}_{D},\vec{\theta}_{G},z)\right)Z\Biggr).\label{eq:QuantumOptimizationTask}
\end{equation}

It is possible to train the circuit of Figure \ref{fig:StructureQGAN}
using gradient descent methods \citep{Salimans2016}. Depending on whether $D(\vec{\theta}_{D}^{k})$
or $G(\vec{\theta}_{G}^{k})$ is being trained at a specific
step $k$, the update rule of the parameters are given by 
\begin{equation}
\begin{array}{rcl}
\vec{\theta}_{D}^{k+1} & = & \vec{\theta}_{D}^{k}+\chi_{D}^{k}\nabla_{\vec{\theta}_{D}}V(\vec{\theta}_{D}^{k},\vec{\theta}_{G}^{k})\\
\\
\vec{\theta}_{G}^{k+1} & = & \vec{\theta}_{G}^{k}-\chi_{G}^{k}\nabla_{\vec{\theta}_{G}}V(\vec{\theta}_{D}^{k},\vec{\theta}_{G}^{k}),
\end{array}\label{eq:UpdateRules}
\end{equation}
where $\chi_{D}^{k}$ and $\chi_{G}^{k}$ are learning rates
which can depend on $k$ in general.

\subsubsection{Limit cases of the training}

The probability that $D(\vec{\theta}_{D})$ successfully
assigns the correct label to $R$ and $G(\vec{\theta}_{G})$
is given by the cost function $V(\vec{\theta}_{D},\vec{\theta}_{G})$.
In what follows we will refer to this probability as $\Pr\left(\mathrm{Success\:}D(\vec{\theta}_{D})|\vec{\theta}_{G}\right)$.
In the ideal case where $G(\vec{\theta}_{G}^{*})=R$, G perfectly
reproduces the statistics of the data source, $D$ cannot distinguish
\citep{Buhrman2001,Aaronson2007} between $R$ and $G(\vec{\theta}_{G}^{*})$,
and $\Pr\left(\mathrm{Success\:}D(\vec{\theta}_{D})|\vec{\theta}_{G}\right)=\frac{1}{2}$.
At this point the training is finished as $D$ cannot improve its
strategy and all gradients vanish:
\begin{equation}
\begin{array}{rcl}
\nabla_{\vec{\theta}_{D}}V(\vec{\theta}_{D},\vec{\theta}_{G}^{*}) & = & 0,\\
\\
\nabla_{\vec{\theta}_{G}}V(\vec{\theta}_{D},\vec{\theta}_{G}^{*}) & = & 0.
\end{array}\label{eq:PerfectTrainingGradients}
\end{equation}
During the training, $\Pr\left(\mathrm{Success\:}D(\vec{\theta}_{D})|\vec{\theta}_{G}\right)$
is bounded by the purity function 
\begin{equation}
C(\vec{\theta}_{G})\equiv\mathrm{tr}\left(\rho^{R}\rho^{G}(\vec{\theta}_{G})\right),\label{eq:Purity}
\end{equation}
such that the performance of the discriminator is
\begin{equation}
\frac{1}{2}C(\vec{\theta}_{G})\leq\Pr\left(\mathrm{Success\:}D(\vec{\theta}_{D})|\vec{\theta}_{G}\right)\leq1-\frac{1}{2}C(\vec{\theta}_{G}).\label{eq:PerformanceDiscriminator}
\end{equation}
The purity function $C(\vec{\theta}_{G})$ is itself bounded
by the nature of $R$. If we define $r_{\min}$ as being the minimal
eigenvalue of $\rho^{R}$, then
\begin{equation}
r_{\min}\leq C(\vec{\theta}_{G})\leq\mathrm{tr}\left
((\rho^{R})^{2}\right),\label{eq:BoundsPurity}
\end{equation}
where the upper bound corresponds to the purity of $\rho^{R}$. 

It is possible to train the circuit of Figure \ref{fig:StructureQGAN}
by evaluating gradients from a numerical finite difference method.
This requires sampling many points around each $(\vec{\theta}_{D},\vec{\theta}_{G})$
to estimate the gradient of \eqref{eq:QuantumOptimizationTask}. In
the following section, we will show how gradients can be evaluated
directly on a quantum computer and explicitly construct the circuits
to optimize \eqref{eq:QuantumOptimizationTask}.

\subsection{Quantum gradients\label{subsec:QuantumGradients}}

\begin{figure}
\begin{centering}
\includegraphics[width=0.98\columnwidth]{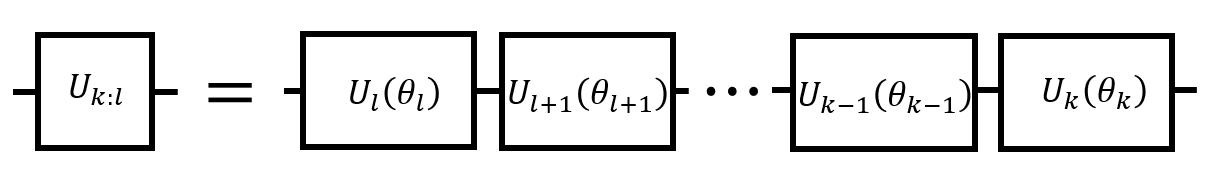}\caption{This notation is used to signify the decomposition of a unitary transformation
in its elementary parametrized gates.\label{fig:UnitaryDecompositionNotation}}
\par\end{centering}
\end{figure}

A key element of GANs is that the generator can be optimized by using gradient signals obtained from the discriminator. Thus, in addition to quantum circuits for $G$ and $D$, we would also like to have quantum circuits which can compute the required gradients. Given access to these quantum gradients, model parameters can be updated via gradient descent on a classical computer.
We introduce some notation useful to define gradient extraction
on a quantum computer \citep{Knill1998,Ortiz2002,Romero2017a,Farhi2018,Schuld2018,Schuld2018a}. In order to present a specific circuit setup, from here onwards we fix that the subsystems of our quantum computer are qubits. We also note that, in addition to the particular setup we use here, there can be other approaches for using a quantum computer to compute gradients of quantum circuits.
A unitary transformation $U$ parametrized by a vector $\vec{\theta}$
with $N$ components is denoted 
\begin{equation}
\begin{array}{rcl}
U(\vec{\theta}) & \equiv & U_{N}\left(\theta_{N}\right)U_{N-1}\left(\theta_{N-1}\right)\ldots U_{2}\left(\theta_{2}\right)U_{1}\left(\theta_{1}\right)\\
\\
 & = & \mathcal{T}\prod_{j=1}^{N}U_{j}\left(\theta_{j}\right),
\end{array}\label{eq:UnitaryDecomposition}
\end{equation}
where $\mathcal{T}$ is the time-ordering operator. It is convenient
to introduce the ordered notation \citep{Machnes2011}
\begin{equation}
\begin{array}{rcl}
U_{k:l} & \equiv & U_{k}\left(\theta_{k}\right)U_{k-1}\left(\theta_{k-1}\right)\ldots U_{l+1}\left(\theta_{l+1}\right)U_{l}\left(\theta_{l}\right)\\
\\
 & = & \mathcal{T}\prod_{j=l}^{k}U_{j}\left(\theta_{j}\right),
\end{array}\label{eq:OrderedNotation}
\end{equation}
which can also be represented in a quantum circuit notation as shown
in Figure \ref{fig:UnitaryDecompositionNotation}. In the same fashion,
the anti-ordered notation has the form
\begin{equation}
\begin{array}{rcl}
U_{l:k}^{\dagger} & \equiv & U_{l}^{\dagger}\left(\theta_{l}\right)U_{l+1}^{\dagger}\left(\theta_{l+1}\right)\ldots U_{k-1}^{\dagger}\left(\theta_{k-1}\right)U_{k}^{\dagger}\left(\theta_{k}\right)\\
\\
 & = & \bar{\mathcal{T}}\prod_{j=l}^{k}U_{j}^{\dagger}\left(\theta_{j}\right),
\end{array}\label{eq:AntiOrderedNotation}
\end{equation}
where $\bar{\mathcal{T}}$ is the anti-time ordering operator. It
follows that we can generally denote $U(\vec{\theta})=U_{N:1}$
and $U^{\dagger}(\vec{\theta})=U_{1:N}^{\dagger}$.

Assuming
each element is generated by a Hamiltonian $h_{j}=h_{j}^{\dagger}$,
an individual gate has the form 
\begin{equation}
U_{j}\left(\theta_{j}\right)=e^{-\frac{i}{2}\theta_{j}h_{j}},\label{eq:UnitaryElementParametrization}
\end{equation}
such that $U_{j}^{\dagger}\left(\theta_{j}\right)=e^{\frac{i}{2}\theta_{j}h_{j}}$.
The derivative of gate $j$ with respect to parameter $\theta_{j}$
is given by 
\begin{equation}
\frac{\partial}{\partial\theta_{j}}U_{j}\left(\theta_{j}\right)=-\frac{i}{2}h_{j}U_{j}\left(\theta_{j}\right).\label{eq:DerivativeUnitaryElement}
\end{equation}
Using the chain rule, we find that 
\begin{equation}
\begin{array}{rcl}
\frac{\partial}{\partial\theta_{j}}U(\vec{\theta}) & = & -\frac{i}{2}U_{N:j+1}h_{j}U_{j:1}\\
\\
\frac{\partial}{\partial\theta_{j}}U^{\dagger}(\vec{\theta}) & = & \frac{i}{2}U_{1:j}^{\dagger}h_{j}U_{j+1:N}^{\dagger}.
\end{array}\label{eq:DerivativeUnitary}
\end{equation}
If we define an initial state on $q$ qubits as $\rho_{0}$,
the expectation value of an observable $P$ evaluated for parameters
$\vec{\theta}$ is given by 
\begin{equation}
\left\langle P(\vec{\theta})\right\rangle =\mathrm{tr}\left(\rho_{0}U^{\dagger}(\vec{\theta})PU(\vec{\theta})\right).\label{eq:ExpectationValue}
\end{equation}
The gradient with respect to a parameter $\theta_{j}$ is then given by
\begin{equation}
\frac{\partial}{\partial\theta_{j}}\left\langle P(\vec{\theta})\right\rangle =-\frac{i}{2}\mathrm{tr}\left(\rho_{0}U_{1:j}^{\dagger}\left[U_{j+1:N}^{\dagger}PU_{N:j+1},h_{j}\right]U_{j:1}\right),\label{eq:GradientExpectationValue}
\end{equation}
where $\left[\cdot,\cdot\right]$ is the commutator. 

At this point it is convenient to introduce some canonical
quantum gates \citep{Nielsen2009}. Specifically, the Hadamard gate
is defined as $H=\frac{1}{\sqrt{2}}\left(\begin{array}{cc}
1 & 1\\
1 & -1
\end{array}\right)$, the NOT gate as $X=\left(\begin{array}{cc}
0 & 1\\
1 & 0
\end{array}\right)$ and $Z=\left(\begin{array}{cc}
1 & 0\\
0 & -1
\end{array}\right)$. It is also useful to define the single-qubit $W$ gate as
\begin{equation}
W\equiv e^{-i\frac{\pi}{4}X}=\frac{1}{\sqrt{2}}\left(\begin{array}{cc}
1 & -i\\
-i & 1
\end{array}\right).\label{eq:WGate}
\end{equation}
\begin{figure}
\begin{centering}
\includegraphics[width=\columnwidth]{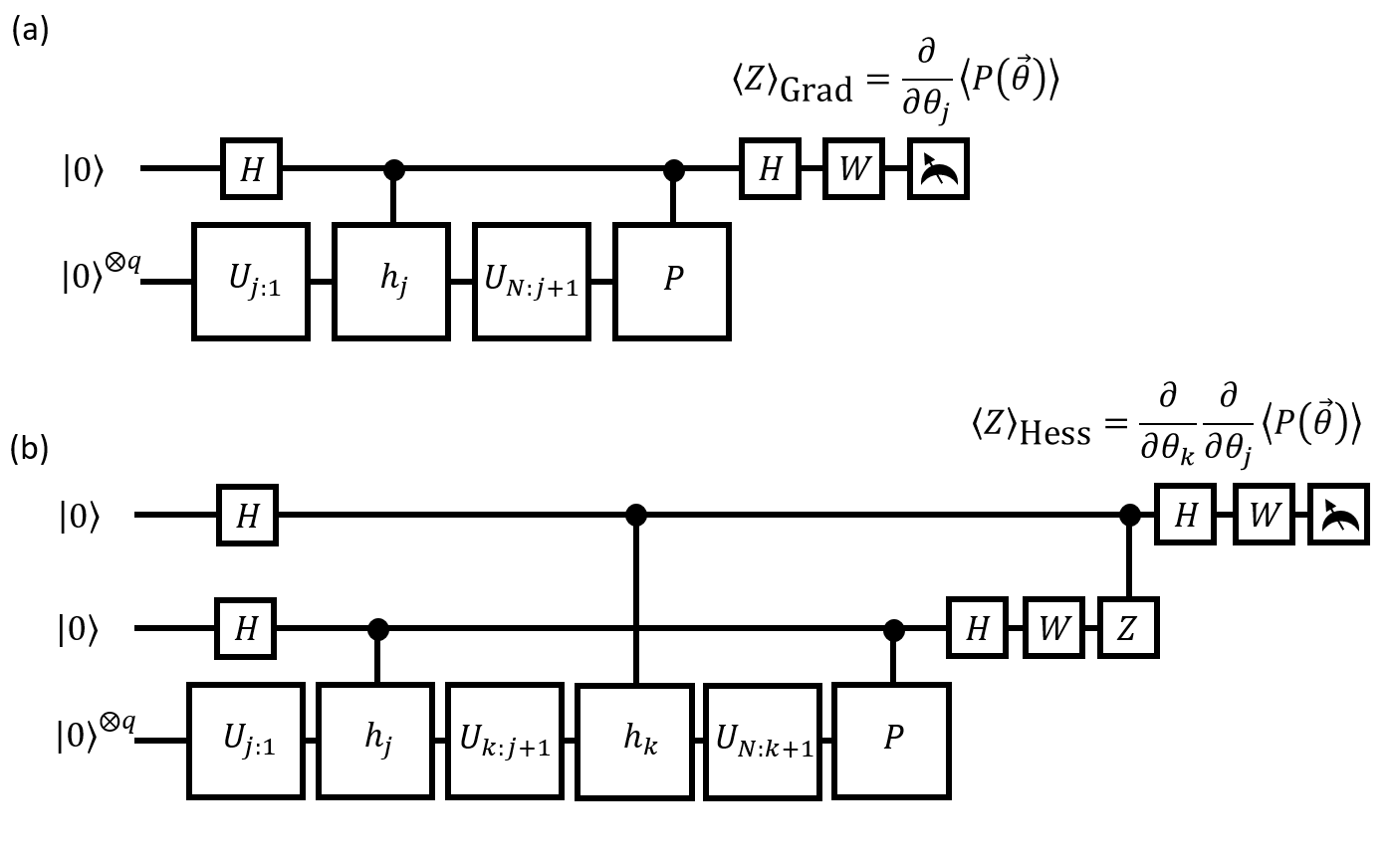}\caption{In (a), we show the general structure of quantum gradients and the structure of quantum Hessians is shown in (b). \label{fig:StructureQuantumGradients}}
\par\end{centering}
\end{figure}
As shown in Figure
\ref{fig:StructureQuantumGradients} (a), the gradient of a parametrized
quantum circuit can be sampled from the $\left\langle Z\right\rangle _{\mathrm{Grad}}$
expectation value of an ancillary qubit such that
\begin{equation}
\begin{array}{rcl}
\left\langle Z\right\rangle _{\mathrm{Grad}} & = & \Pr\left(\left|x_{\mathrm{Grad}}\right\rangle =\left|0\right\rangle \right)-\Pr\left(\left|x_{\mathrm{Grad}}\right\rangle =\left|1\right\rangle \right)\\
\\
 & = & \frac{\partial}{\partial\theta_{j}}\left\langle P(\vec{\theta})\right\rangle .
\end{array}\label{eq:GradientQuantumCircuit}
\end{equation}
Note that this requires the ability to perform control gates for the Hamiltonians $h_j$ and measurement operator $P$.
Similarly, using the fact that the Hessian is the gradient of a gradient,
we show how the Hessian can be measured in Figure \ref{fig:StructureQuantumGradients} (b),
such that the output is
\begin{equation}
\left\langle Z\right\rangle _{\mathrm{Hess}}=\frac{\partial^{2}}{\partial\theta_{k}\partial\theta_{j}}\left\langle P(\vec{\theta})\right\rangle .\label{eq:HessianQuantumCircuit}
\end{equation}

\subsection{Using quantum gradients to train QuGANs}
\label{subsec:GradientsQGANs}

\begin{figure}
\centering{}\includegraphics[width=\columnwidth]{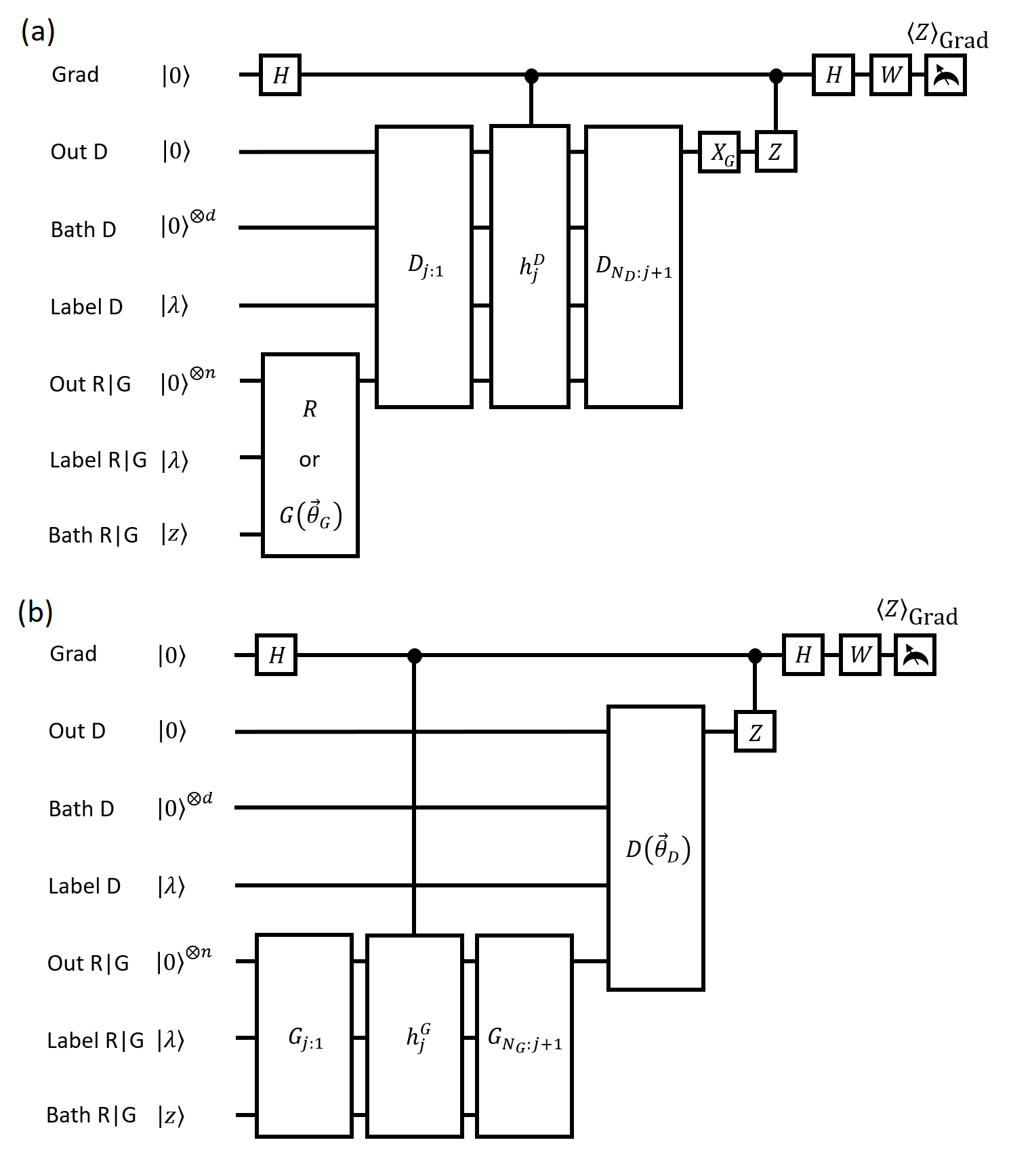}\caption{In (a), we show the quantum circuit used to measure gradient of the discriminator with real data (using $X_G=I$ and fixing $\left|z=0\right\rangle$) and with generated data (using $X_G=X$). The circuit to measure the gradient of the generator is shown in (b).\label{fig:GradientDG}}
\end{figure}

We now have all the elements required to evaluate the gradients of
\eqref{eq:UpdateRules} directly on a quantum computer. The operator
$P$ from Section \ref{subsec:QuantumGradients} corresponds to the
$Z$ operator of \eqref{eq:DecisionDipole} when computing gradients.
The parametrized discriminator $D$ and generator $G$ can be decomposed
into respectively $N_{D}$ and $N_{G}$ gates, such that
\begin{equation}
\begin{array}{rcl}
D(\vec{\theta}_{D}) & = & D_{N_{D}:1},\\
\\
G(\vec{\theta}_{G}) & = & G_{N_{G}:1}.
\end{array}\label{eq:GateDecompositionDG}
\end{equation}
In order to measure gradients, we introduce a single-qubit register
\textbf{Grad}. It follows that all elements $\frac{\partial}{\partial\theta_{Dj}}V(\vec{\theta}_{D},\vec{\theta}_{G})=\frac{1}{4\Lambda}\left\langle Z\right\rangle _{\mathrm{Grad}}$
of the gradient of the discriminator 
\begin{multline}
\frac{\partial}{\partial\theta_{Dj}}V(\vec{\theta}_{D},\vec{\theta}_{G})=-\frac{i}{8\Lambda}\sum_{\lambda=1}^{\Lambda}\mathrm{tr}\Biggl(\left(\rho_{\lambda}^{R}-\rho_{\lambda}^{G}(\vec{\theta}_{G},z)\right)\\
\times U_{D,1:j}^{\dagger}\left[U_{D,j+1:N_{D}}^{\dagger}ZU_{D,N_{D}:j+1},h_{j}^{D}\right]U_{D,j:1}\Biggr),\label{eq:GradientDiscriminator}
\end{multline}
can be evaluated for each label and sources $R$ and $G(\vec{\theta}_{G})$ by the quantum circuit of Figure \ref{fig:GradientDG} (a) with an appropriate $X$ gate to account for the sign of the cost function. In the later
case, an $X$ gate is applied on the \textbf{Out D} register after
the discriminator to get the correct sign of the gradient. The circuit
that yields the gradient 
\begin{multline}
\frac{\partial}{\partial\theta_{Gj}}V(\vec{\theta}_{D},\vec{\theta}_{G})=\frac{i}{8\Lambda}\sum_{\lambda=1}^{\Lambda}\mathrm{tr}\Biggl(\rho_{\lambda}^{0}(z)\\
\times U_{G,1:j}^{\dagger}\left[U_{G,j+1:N_{G}}^{\dagger}U_{D}^{\dagger}(\vec{\theta}_{D})ZU_{D}(\vec{\theta}_{D})U_{G,N_{G}:j+1},h_{j}^{G}\right]U_{G,j:1}\Biggr)\label{eq:GradientGenerator}
\end{multline}
of the generator $-\frac{\partial}{\partial\theta_{Gj}}V(\vec{\theta}_{D},\vec{\theta}_{G})=\frac{1}{4\Lambda}\left\langle Z\right\rangle _{\mathrm{Grad}}$
for each label is shown in Figure \ref{fig:GradientDG} (b). We note that the
sign is meant to be the same as the one in \eqref{eq:UpdateRules},
such that the generator improves its capability to fool the discriminator.
More advanced methods to update the parameters could also leverage
the use of quantum Hessians \eqref{eq:HessianQuantumCircuit}.

\subsubsection{Improved training heuristics\label{subsec:ImprovedTrainingHeuristics}}

Training GANs is equivalent to finding the Nash equilibrium of a two-player
game. This problem is known to be in the complexity class PPAD which
is not expected to be contained in BQP \citep{Fellman2010,Li2011}.
Advanced heuristics have been developed to improve the training of
classical GANs \citep{Salimans2016}. Namely, it should be straightforward
to implement semi-supervised learning in the quantum context by increasing
the number of labels to $\Lambda+1$ and supplying some labeled examples
of generated data. Feature matching should also be possible by truncating
the decomposition of $D(\vec{\theta}_{D})$ when evaluating
the gradients of $G(\vec{\theta}_{G})$ with the circuit
of Figure \ref{fig:GradientDG} (b). We also assumed that the expectation
value of each gradient is evaluated from ensemble averaging; it may also
be possible to use Bayesian methods to update the parameters after
single-shot measurements \citep{Stenberg2016}.

\subsection{A practical ansatz\label{subsec:Ansatz}}

\begin{figure*}
\centering{}\includegraphics[width=0.8\textwidth]{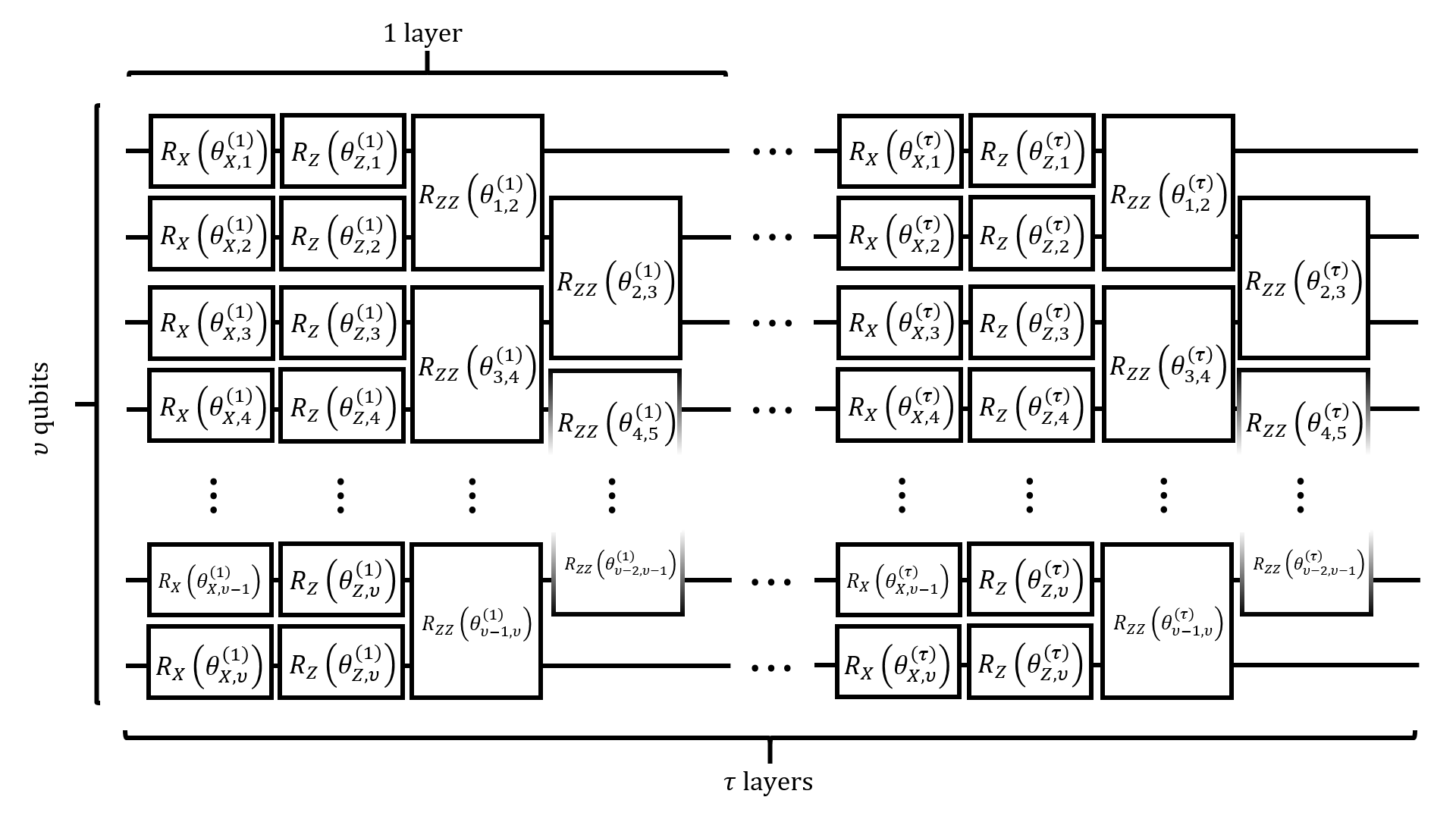}\caption{A practical circuit ansatz for the generator $G$ and the discriminator
$D$ composed of $\tau$ layers acting on $\nu$ qubits. Each layer
$t$ is composed of single-qubit $X$ rotations parametrized by angles
$\vec{\theta}_{X}^{\left(t\right)}=\left\{ \theta_{X,1}^{\left(t\right)},\ldots,\theta_{X,\nu}^{\left(t\right)}\right\} $
followed by $Z$ rotations parametrized by $\vec{\theta}_{Z}^{\left(t\right)}=\left\{ \theta_{Z,1}^{\left(t\right)},\ldots,\theta_{Z,\nu}^{\left(t\right)}\right\} .$
A layer of two staggered sets of nearest-neighbor $ZZ$ rotations
parametrized by $\vec{\theta}_{ZZ}^{\left(t\right)}=\left\{ \theta_{1,2}^{\left(t\right)},\ldots,\theta_{\nu-1,\nu}^{\left(t\right)}\right\} $
follows the single-qubit rotations. The ansatz is universal for quantum
computing in the limit of an infinite number of layers since it can
generate arbitrary single-qubit gates as well as entangling two-qubit
gates. \label{fig:CircuitAnsatz}}
\end{figure*}

A potentially useful ansatz to parametrize $D(\vec{\theta}_{D})$
and $G(\vec{\theta}_{G})$ is shown in Figure \ref{fig:CircuitAnsatz}.
It is universal for quantum computing in the limit of an infinite
number of layers $\tau$. Since the generators of those gates are
all simple Pauli operators, it is easy to implement the conditional
$h_{j}$'s with CNOTs, CPHASEs and CZZs where the $ZZ$s are between
nearest-neighbor qubits. Other types of ansatz may be used depending
on the context \citep{Peruzzo2014,Kandala2017,Dallaire-Demers2018,Schuld2018a,Huggins2018}.

\subsection{Numerics\label{subsec:Numerics}}

We numerically tested ideas in this paper with a simple example involving two labels $A$ and $B$. We chose a source $R$ such that $\rho^R_A=\left|0\right\rangle\left\langle 0\right|$ and $\rho^R_B=\left|1\right\rangle\left\langle 1\right|$. The labels can be encoded in a 1-qubit \textbf{Label R|G} register and the \textbf{Out R|G} register only requires 1 qubit. Since the labeled distributions $\rho^R_{\lambda}$ are pure we don't need a \textbf{Bath R|G} register to generate entropy. The expected solution is that $G$ should be able to generate a CNOT gate conditioned on the label register. We find that this can be achieved with 2 layers of the ansatz previously introduced. This corresponds to 10 variational parameters in  $\vec{\theta}_{G}$.

The discriminator requires at least 1 qubit for its output \textbf{Out D}, 1 qubit for \textbf{Label D}, and it also operates on the 1-qubit \textbf{Out R|G} register. We find that a \textbf{Bath D} register did not appear to improve convergence of our numerical experiments. Therefore, $D$ operates on 3 qubits, and we found that 4 layers of the ansatz of section \ref{subsec:Ansatz} were sufficient to train the QuGAN. This yields 32 parameters in $\vec{\theta}_{D}$ for a total of 42. With the qubit of register \textbf{Grad}, the algorithm operates on a total of 5 qubits.

Training GANs is a delicate art. To keep this proof-of-principle simple we chose not to use any advanced training heuristic. We trained the QuGAN for 10,000 gradient steps of the update rule \eqref{eq:UpdateRules}. The learning rate $\chi_D^k$ exponentially decreases from 10 to $\frac{1}{10}$ for the first 4,000 steps and remains constant at the latter value for the remaining 6,000 steps. The generator $G$ is only updated once for every 100 steps of $D$ with a learning rate $\chi_G^k=5\chi_D^k$.

\begin{figure}
\centering{}\includegraphics[width=0.9\columnwidth]{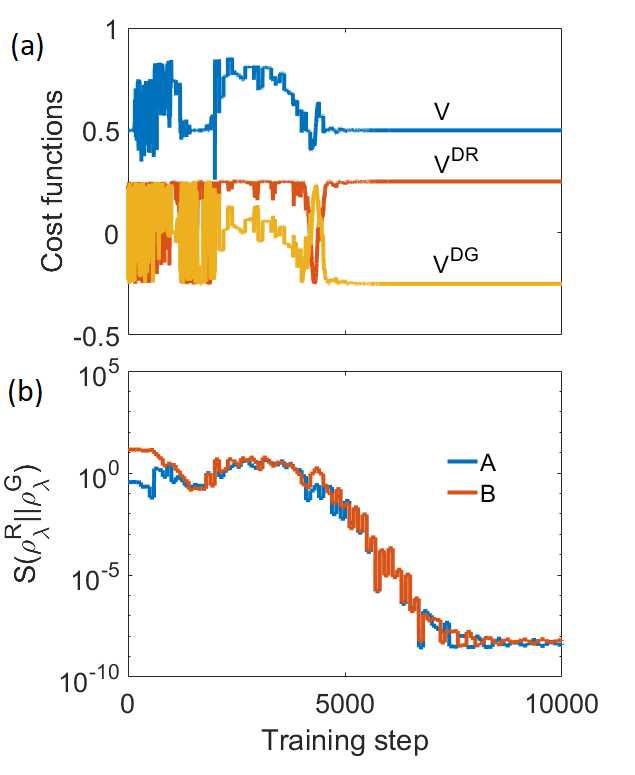}\caption{ A source $R$ produces two labeled states $\left|A,0\right\rangle$ and $\left|B,1\right\rangle$.
In (a) we have the values of the cost functions as a function of the training step.
In (b) we show the cross-entropy $S\left(\rho^R_{\lambda}\|\rho^G_{\lambda}\right)$ for each labeled distribution as a function of the training step.
\label{fig:Numerics}}
\end{figure}

As shown in Figure \ref{fig:Numerics}, the generator has been properly trained at the end of the algorithm, as the cross-entropy
\begin{equation}
S\left(\rho^R_{\lambda}\|\rho^G_{\lambda}\right) = \mathrm{tr}\left(\rho^R_{\lambda}\left(\mathrm{log}_2\rho^R_{\lambda}-\mathrm{log}_2\rho^G_{\lambda}\right)\right)
\label{eq:CrossEntropy}
\end{equation}
quickly converges to zero. We also plotted the components of the cost function $V$ defined as
\begin{equation}
\begin{array}{rcl}
V^{DR}(\vec{\theta}_{D})&=&\frac{1}{4\Lambda}\sum_{\lambda=1}^{\Lambda}\mathrm{tr}\left(\rho_{\lambda}^{DR}(\vec{\theta}_{D})Z\right)\\
\\
V^{DG}(\vec{\theta}_{D},\vec{\theta}_{G})&=&-\frac{1}{4\Lambda}\sum_{\lambda=1}^{\Lambda}\mathrm{tr}\left(\rho_{\lambda}^{DG}(\vec{\theta}_{D},\vec{\theta}_{G},z)Z\right),
\end{array}\label{eq:CostFunctionComponents}
\end{equation}
such that $V=\frac{1}{2}+V^{DR}+V^{DG}$. At the beginning of the training sequence, the parameters are chosen randomly. In order to provide a reliable training signal for $G$, the gradients are amplified by a large learning rate to quickly train $D$. The generator initially produces a decent state for the $A$ label but fails to produce a good state for the $B$ label. The training of the discriminator appears successful since $V$ is typically larger than $\frac{1}{2}$ and learns to differentiate the data produced by $R$ from the data produced by $G$. Updating the generator less often than the discriminator provides a trade-off between a fast training of $G$ and a good training signal. After a few tens of training steps of $G$ (which corresponds to a few thousand training cycles of $D$) the cross-entropy between the real and the generated data starts to converge to zero as the generator creates better samples. In this case, $D$ cannot differentiate $R$ and $G$ as $V$ approaches its equilibrium value of $\frac{1}{2}$. The final strategy of $D$ is to designate all data as real, yielding $V^{DR}\approx\frac{1}{2}$ and $V^{DG}\approx-\frac{1}{2}$.\\

\section{Conclusion\label{sec:Conclusion}}

Quantum machine learning is likely to be one of the first general-purpose
applications of near-term quantum devices. Here we showed how generative
models can be trained on quantum computers. We have reformulated the
optimization problem of GANs in the quantum formalism, yielding QuGANs.
We have shown how the cost function can be optimized by directly evaluating
the gradients with a quantum processor. We provided a simple
universal qubit ansatz which constrains the set of additional quantum resources
required to evaluate the gradients. Finally, we showed that QuGANs can be trained in practice by performing a simple numerical experiment.

It is expected that QuGANs will have a more versatile representation
power than their classical counterpart. For example, one can speculate
that a large enough QuGAN could learn to generate encrypted data labeled
by RSA public encryption keys since quantum computers have the capacity to perform Shor factoring \citep{Shor1997}
and hence decryption. 
In that case, the optimal generator would learn a statistical model
of the unencrypted data for each key and encrypt with the label. Other
classical cryptographic systems (such as elliptic curve) could also be vulnerable to this type of attack. 
In this work, we have explored the practical issues of QuGANs, namely, explicit quantum circuits for the generator and discriminator, as well as quantum methods for computing the gradients of these circuits. 
A more general analysis of the theoretical concepts of quantum adversarial learning can be found in the companion paper \cite{lloyd2018}. 
\begin{acknowledgments}
We thank Seth Lloyd and Christian Weedbrook for their insightful advices. This works was made possible by ample supplies of Tim Horton's coffee. 
\end{acknowledgments}

\bibliographystyle{apsrev4-1}
\bibliography{QGAN}

\end{document}